\documentclass[aps,prl,twocolumn,groupedaddress,superscriptaddress,showpacs]{revtex4}

\usepackage{graphicx}
\usepackage{dcolumn}
\usepackage{bm}
\usepackage[usenames]{color}
\usepackage{ulem}

\begin{document}

\title{Spin-helical Dirac states in graphene induced by polar-substrate surfaces with giant spin-orbit interaction: a new platform for spintronics}

\author{S.~V. Eremeev}
 \affiliation{Institute of Strength Physics and Materials Science, 634021, Tomsk, Russia\\}
 \affiliation{Tomsk State University, 634050, Tomsk, Russia\\}
 \affiliation{Donostia International Physics Center (DIPC), 20018 San Sebasti\'an/Donostia, Basque Country, Spain\\}

\author{I.~A. Nechaev}
 \affiliation{Tomsk State University, 634050, Tomsk, Russia\\}
 \affiliation{Donostia International Physics Center (DIPC), 20018 San Sebasti\'an/Donostia, Basque Country, Spain\\}

\author{P.~M. Echenique}
 \affiliation{Donostia International Physics Center (DIPC), 20018 San Sebasti\'an/Donostia, Basque Country, Spain\\}
 \affiliation{Departamento de F\'{\i}sica de Materiales UPV/EHU, Facultad de Ciencias Qu\'{\i}micas, UPV/EHU, Apdo. 1072, 20080 Sebasti\'an/Donostia, Basque Country, Spain\\}
 \affiliation{Centro de F\'{\i}sica de Materiales CFM - MPC, Centro Mixto CSIC-UPV/EHU, 20080 San Sebasti\'an/Donostia, Basque Country, Spain\\}

\author{E.~V. Chulkov}
 \affiliation{Donostia International Physics Center (DIPC), 20018 San Sebasti\'an/Donostia, Basque Country, Spain\\}
 \affiliation{Departamento de F\'{\i}sica de Materiales UPV/EHU, Facultad de Ciencias Qu\'{\i}micas, UPV/EHU, Apdo. 1072, 20080 Sebasti\'an/Donostia, Basque Country, Spain\\}
 \affiliation{Centro de F\'{\i}sica de Materiales CFM - MPC, Centro Mixto CSIC-UPV/EHU, 20080 San Sebasti\'an/Donostia, Basque Country, Spain\\}

\date{\today}

\begin{abstract}
Spintronics, or spin electronics, is aimed at efficient control and
manipulation of spin degrees of freedom in electron systems. To
comply with demands of nowaday spintronics, the studies of
electron systems hosting giant spin-orbit-split electron states have
become one of the most important directions providing us with a
basis for desirable spintronics devices. In construction of such
devices, it is also tempting to involve graphene, which has attracted
great attention because of its unique and remarkable electronic
properties and was recognized as a viable replacement for silicon in
electronics. In this case, a challenging goal is to make graphene
Dirac states spin-polarized. Here, we report on absolutely new
promising pathway to create spin-polarized Dirac states based on
coupling of graphene and polar-substrate surface states with giant
Rashba-type spin-splitting. We demonstrate how the spin-helical
Dirac states are formed in graphene deposited on the surface of
BiTeCl. This coupling induces spin separation of the originally
spin-degenerate graphene states and results in fully helical
in-plane spin polarization of the Dirac electrons.
\end{abstract}


\maketitle

\section{Introduction}

Graphene is a fascinating material, which has attracted great
attention because of its unique electronic properties
\cite{GN_NatMat2007,Castro_RevModPhys2009}. In graphene spintronics,
many efforts were made to realize a robust control of electron spins
by, e.g., magnetoelectric coupling or spin-orbit interaction (SOI)
\cite{Zutic_RevModPhys2004,Awschalom_NatPhys2007,Zhang_PRL2009}.
Main hopes were pinned on the SOI effect, which can be directly
observed \cite{Kato_Sci2004,Wunderlich_PRL2005}.

Some interesting phenomena, such as quantum spin Hall effect
\cite{KaneMele_PRL2005}, quantum anomalous Hall effect
\cite{Qiao_PRB2010}, and other phenomena were predicted in graphene.
However, the intrinsic spin-orbit splitting in pristine graphene is
proved to be too weak to produce an observable effect and to realize
practical applications
\cite{KaneMele_PRL2005,Yao_PRB2007,Gmitra_PRB2009}. The major
challenge in graphene spintronics is to make graphene Dirac states
spin-polarized. In principle, it can be realized by applying an
extrinsic SOI, i.e., by placing the grapnene in a proper medium with
a strong spin-orbit coupling. Numerous previous works, both
experimental and theoretical,  have been aimed at enhancement of
graphene SOI via adatoms deposition
\cite{Castro_PRL2009,Abdelouahed_PRB2010,Qiao_PRB2010,Weeks_PRX} or
growth of graphene on metal substrates
\cite{Dedkov_PRL2008,Varykhalov_PRL2008,Heer_JPhD2010,Li_JPCM2011,Marchenko_NatCom2012}.

In the present paper, we propose a new pathway to manipulate the
electronic and spin properties of graphene by its depositing on a
polar-substrate surface possessing a giant Rashba-type spin-split
surface state. We state that one of the efficient mechanisms
responsible for appearance of spin-polarised Dirac states in
graphene is the coupling of two two-dimensional (2D) electron
systems: Dirac electrons of graphene and Rashba electrons of the
mentioned surface state. To corroborate our statement, we have
chosen the most suitable 2D Rashba-electron system formed by
electons in the surface state at the Te-terminated surface of
bismuth tellurohalides. Here, on the base of relativistic density
functional theory (DFT) calculations, we show that graphene
deposited on BiTeCl is the system with unique electronic properties,
in which the strong hybridization of Dirac and  Rashba electrons
gives rise to spin-helical Dirac states.

\section{Method}

Our calculations were based on DFT as implemented in the Vienna {\it
ab initio} simulation package VASP, \cite{VASP1,VASP2}. The
generalized gradient  approximation (GGA) of Perdew, Burke, and
Ernzerhof (PBE) \cite{GGA_PBE} to the exchange correlation (XC)
potential has been implemented. The interaction  between the ion
cores and valence electrons was described by the projector
augmented-wave method \cite{PAW1,PAW2}. The Hamiltonian contained
the scalar relativistic corrections, and the spin-orbit coupling was
taken into account by the second variation method \cite{Koelling}.
To simulate BiTeCl substrate we  consider a 24 atomic layer slab.
Hydrogen monolayer was used to passivate the chlorine side of the
slab as was described in Ref.~\cite{Eremeev_PRL}. The  positions of
atoms in the graphene layer and within three outermost atomic layers
of the BiTeCl slab were optimized including SOI self-consistently.
The distance between the substrate and the graphene layer was found
to be of $\approx 3.35$\AA. The atoms of the deeper layers were kept
fixed at the bulk crystalline positions. The $k$-point mesh of
$9\times 9\times 1$ was used for the Brillouin zone of the surface
unit cell.

\begin{figure}
\includegraphics[width=\columnwidth]{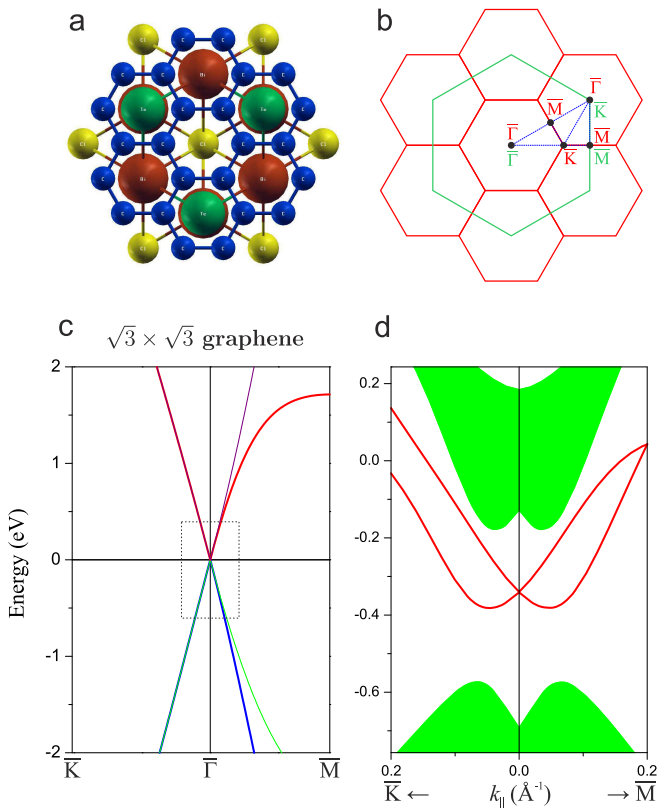}
\caption{(Color online) {\bf Atomic structure of graphene on BiTeCl
and electronic spectra of the pristine graphene and Te-terminated
BiTeCl surface.} ({\bf a}) Top view for the $\sqrt{3}\times
\sqrt{3}$ graphene on the Te-terminated surface of BiTeCl.
Carbon atoms are shown by dark blue balls; green, maroon, and yellow
balls denote first-layer Te atom, second-layer Bi atom, and
third-layer Cl atom, respectively; ({\bf b}) Scheme of
$\sqrt{3}\times \sqrt{3}$ folding of the 2D BZ of graphene: green
and red colors correspond to the $1\times 1$ and $\sqrt{3}\times
\sqrt{3}$ BZs, respectively; ({\bf c}) Band structure of the
$\sqrt{3}\times \sqrt{3}$ graphene; ({\bf d}) Electronic spectrum of
the Te-terminated BiTeCl surface. The Rashba-split surface state is
shown by red line, projected bulk band structure is depicted by
green areas.}
 \label{fig1}
\end{figure}

\begin{figure}
\includegraphics[width=\columnwidth]{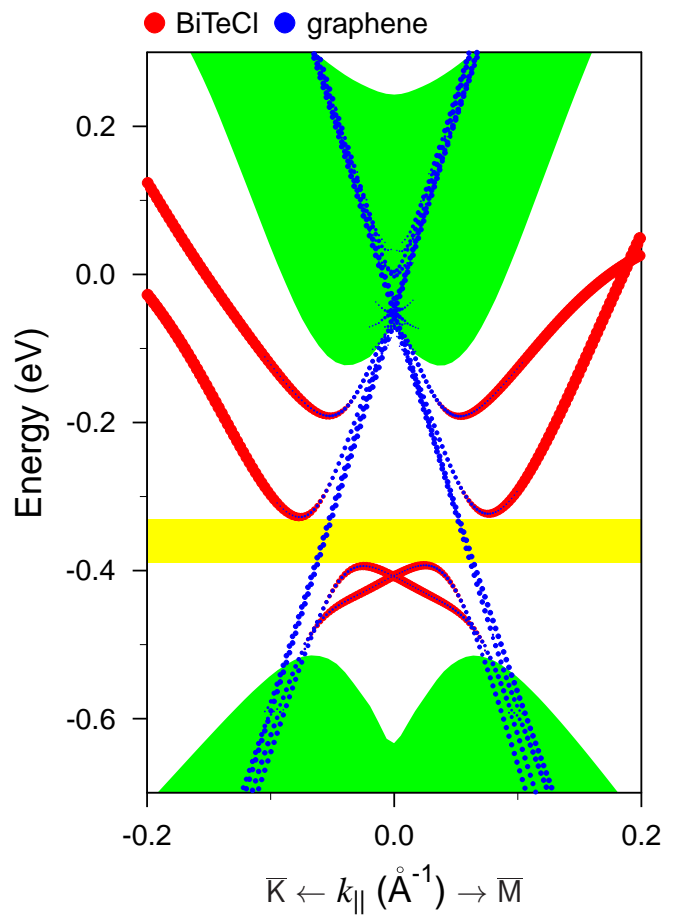}
\caption{(Color online) {\bf Electronic structure of
graphene@BiTeCl.} Band structure of BiTeCl(0001) slab with
$\sqrt{3}\times \sqrt{3}$ graphene on Te-terminated surface. Red and
blue circles denotes weights of the states localized in the
outermost three layers (1-st TL) of BiTeCl and in graphene layer,
respectively. Shaded by green color regions indicate projected bulk
bands of substrate.}
 \label{fig2}
\end{figure}

\begin{figure*}
\includegraphics[width=\textwidth]{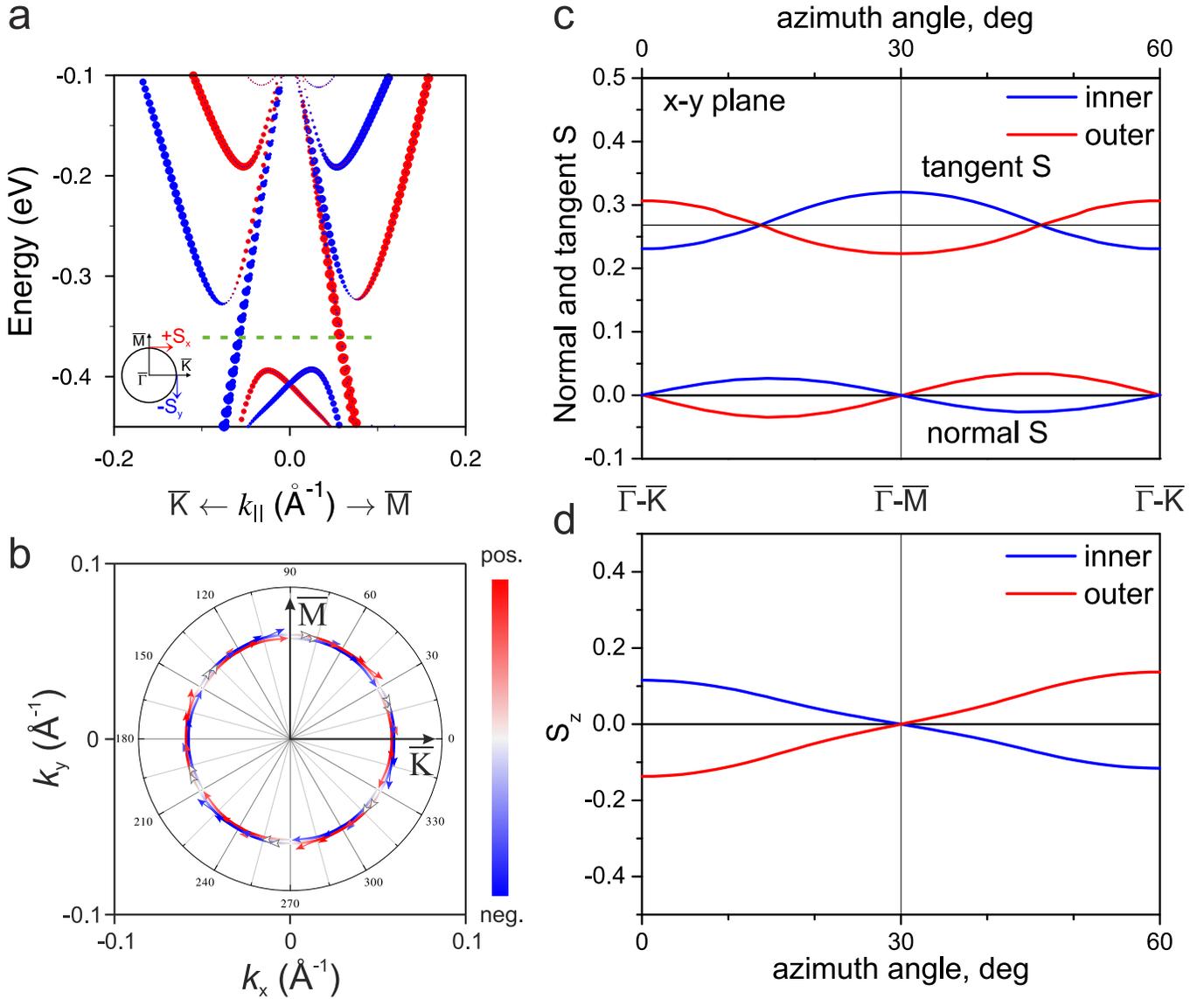}
\caption{(Color online) {\bf Spin structure of the surface states.}
({\bf a}) The spin-resolved electronic spectrum of the surface
states where the filled circles represent the weight of the states
multiplied by the value of the in-plane spin component (red and blue
colors denote positive and negative values of spin,  respectively,
for mutually perpendicular $\bar\Gamma - \bar{\rm M}$ and
$\bar\Gamma - \bar{\rm K}$ directions as schematically shown in
inset); ({\bf b})  Constant energy contours for graphene Dirac-cone
bands at energy of -365 meV, marked by dashed green line in panel
(a). Arrows adjacent to the contours denote the in-plane spin
component. The out-of-plane spin component is indicated by the
colour with red and blue corresponding to the upward and downward
directions, respectively. White colour arrows indicate fully
in-plane spin alignment; Azimuthal dependencies of the graphene spin
components, including the in-plane  components in the tangent and
normal directions to the contours ({\bf c}) and out-of-plane
component ({\bf d}), inner and outer Dirac contours shown by blue
and red, respectively.}
 \label{fig4}
\end{figure*}

\section{Results}

Recently, it was shown that the Te-terminated surface of
non-centrosymmetric hexagonal-structured polar-semiconductor
compounds BiTe(Cl,Br,I) meet requirements for spintronics
applications, since these systems hold a giant Rashba spin splitting
of a free-electron-like surface state at the $\bar \Gamma$ point
\cite{Eremeev_PRL,Crepaldi_PRL,Landolt_PRL,Eremeev_JETPlett,Sakano,Eremeev_NJP}.
The mentioned bismuth tellurohalides are characterized by ionic
bonding with large charge transfer from bismuth to halide- and
tellutium-atomic layers. The spin-split surface states at the
Te-terminated surface emerge by splitting off from the lowest
conduction band, owing to the decreasing potential within the
near-surface layers \cite{Eremeev_PRL}, which is a consequence of
strong ionicity. One can expect that deposition of graphene on
BiTe(Cl,Br,I), can result in a strong interaction and hybridization
of the Rashba and Dirac 2D electrons. In designing such systems, it
is important to match the lattice parameters of the contacting
materials. The in-plane hexagonal parameter of BiTeCl matches
perfectly with the parameter of $\sqrt 3\times \sqrt 3$ graphene.
Thus, the deposited graphene should not undergo the in-plane strain
[Fig.~\ref{fig1}(a)]. In the $\sqrt 3\times \sqrt 3$ structure, the
Brillouin zone of graphene takes a third of the original one
[Fig.~\ref{fig1}(b)]. Under this folding, the $\bar{\rm K}$-point
Dirac cone of graphene appears as four-fold degenerate $\bar\Gamma$
state  [Fig.~\ref{fig1}(c)] with linear dispersion in the vicinity
of the Dirac point (Fig.~\ref{fig1}(c), the dotted-line rectangle).
Thus, both the Dirac state and the Rashba state of the pristine
substrate [Fig.~\ref{fig1}(d)] reside in the Brillouin zone center.

The interaction of the two 2D electron systems results in a strong
modification of the spectra with respect to the constituents
(Fig.~\ref{fig2}). In the  upper part of the gap, at large $k_\|$
the Rashba states of BiTeCl are preserved, while at $k_\| \lesssim
0.1$ \AA$^{-1}$ both inner and outer branches of the Rashba states
rapidly become strongly hybridized with graphene cone states, and
already at $k_\| \approx 0.05$ \AA$^{-1}$ they are completely
localized within the graphene layer, dispersing towards the Dirac
point, which is immersed in the conduction band of the substrate
bulk states. In the lower part of the gap, two of four graphene
bands are hybridized with the substrate forming hole-like Rashba
state near the valence band, which is completely localized within
substrate layers in the vicinity of the $\bar \Gamma$ point. Thus,
hybridization of two graphene branches with the Rashba state of the
BiTeCl substrate leads to a break of both branches of the Rashba
states at small $k_\|$ in such a way that its degeneracy point
appears near the valence band or, to the contrary, the Rashba states
break two graphene branches. Other two Dirac branches remain
unchanged and localized in the graphene layer for all $k_\|$. As a
result of these changes, in the middle part of the bulk band gap
within the energy interval of 70 meV (see yellow stripe in
Fig.~\ref{fig2}) two almost degenerate Dirac states of graphene
survive. The small SOI-induced $k_\|$-splitting for the Dirac cones
is less then 0.002\,\AA$^{-1}$.

The spin structure of the surface states within the $\bar \Gamma$
band gap is shown in Fig.~\ref{fig4}(a). As seen in the figure, at
large $k_\|$ the Rashba branches preserve the spin polarization of
the clean Te-terminated BiTeCl surface: they demonstrate
counter-clockwise and clockwise in-plane helicity for the inner and
outer branches, respectively. At $k_\| < 0.1$ \AA$^{-1}$, where due
to hybridization the Rashba branches turn into the Dirac states and
start to be localized at the graphene layer, the spin helicity of
the graphene band hybridizing with the outer branch becomes the same
as that of the inner branch. Consequently, both hybridized graphene
bands have the counter-clockwise spin polarization. It certainly
would mean that two remaining non-hybridized Dirac bands should be
characterized by the clockwise spin polarization, as it is confirmed
by the spin-resolved bands presented in Fig.~\ref{fig4}(a). Thus,
and this is the main issue of the present study, the hybridization
between the graphene and substrate-Rashba-split 2D electrons
provides the helical spin separation of the graphene Dirac states.

As was mentioned above, the most interesting feature of the
graphene@BiTeCl spectrum is the energy window of 70 meV width, where
the Dirac states of graphene exist only (see yellow stripe in
Fig.~\ref{fig2}). The two almost degenerate Dirac states have the
same in-plane spin polarization, which can be clearly seen in
Fig.~\ref{fig4}(b), where spin-resolved constant energy contours in
the middle of this energy window are shown. Both Dirac states, apart
from the in-plane clockwise spin polarization, also demonstrate the
presence of the out-of-plane spin component, which is intrinsic
feature of the  spin-polarized states at hexagonal surfaces. The
detailed spin texture is illustrated in Figs.~\ref{fig4}(c) and (d).
Fig.~\ref{fig4}(c) shows the azimuthal dependencies of the in-plane
spin components in the tangent and normal directions to the inner
and outer Dirac contours. One can see that the tangent component for
both bands demonstrates variation around the value of 0.27, while
the normal component experiences small variations near zero spin
value. For both bands, the out-of-plane spin component $S_z$
[Fig.~\ref{fig4}(d)] also varies near zero but the variations are
significantly larger than those for the normal component.
Nevertheless, the  maximal out-of-plane spin values, which are
observed along the $\bar\Gamma-\bar{\rm K}$ directions, are three
times smaller than the magnitude of the tangent in-plane spin
component. Furthermore, the signs of both normal and out-of-plane
spin components are opposite for the inner and outer almost
degenerate Dirac contours, what results in a fully helical net
in-plane spin polarization of the Dirac electrons.

\section{Discussion}

Apparently, the revealed possibility to create the helical
spin-polarized Dirac electrons in graphene opens up new horizons in
the graphene-based spintronics. The mentioned spin texture of the
graphene Dirac states is important to come out as a consequence of
the strong hybridization between graphene and giant Rashba-split
states of the Te-terminated surface of the polar substrate BiTeCl.
Due to excellent matching lattice parameters of BiTeCl and graphene,
the graphene@BiTeCl system can be easily realized on practice. The
reported energy interval of $\sim70$ meV in the middle part of the
projected bulk energy gap, where the spin-polarized Dirac states
solely exist, is sufficiently enough for spintronics applications.
However, it should be noted that this energy interval can be wider
in a real system, since the band gap is normally underestimated
within the DFT. As shown both by experiment \cite{Sakano} and by the
$GW$ bulk calculations \cite{Rusinov}, the bulk band gap in BiTeCl
is a factor of two larger than that predicted by the DFT. In order
to actively bring the spin-polarized Dirac states located in the
middle part of the gap into play, the chemical potential of the
system should be decreased by $\sim 350$ meV with respect to the
Fermi level obtained in the present calculation. The chemical
potential tuning can be realized by applying electric field or
doping hole-donating adsorbates. The latter approach seems to be
preferable. As was recently shown for Au/Si system, supporting
Rashba-split surface state, the Fermi level position can be
effectively tuned within the range of $\sim 350$ meV by choosing
appropriate adsorbate species \cite{Bondarenko}.

Except the spin-polarized Dirac states in the middle part of the
gap, the surface states in upper and lower parts of the bulk gap may
also be of considerable  interest, because in the respective energy
regions the helical Dirac states of graphene and Rashba spin-split
states coexists. Various intriguing physical phenomena related to
decay of elementary excitations in the states with different
velocities can be expected.

We believe that our findings will stimulate further theoretical and
experimental investigations of spin-polarized Dirac fermions which
is a key matter for  prospective spintronics applications.

\section{acknowledgements}
We acknowledge partial support from the Basque Country Government,
Departamento de Educaci\'{o}n, Universidades e Investigaci\'{o}n
(Grant No. IT-756-13), the Spanish Ministerio de Ciencia e
Innovaci\'{o}n (Grant No. FIS2010-19609-C02-01), and the Ministry of
Education and Science of Russian Federation (Grant No. 2.8575.2013).


\end{document}